\documentclass[12pt,a4paper]{article}

\usepackage{natbib,graphicx}

\title
{\bf On reverberation and cross-correlation estimates of the size of
the broad-line region in active galactic nuclei}

\author
{A. V. Melnikov\thanks{E-mail: melnikov@gao.spb.ru}\, and I. I. Shevchenko \\
Pulkovo Observatory of the Russian Academy of Sciences,\\
Pulkovskoje ave. 65/1, St.Petersburg 196140, Russia}

\begin{document}

\maketitle

\label{firstpage}

\begin{abstract}
It is known that the dependence of the emission-line luminosity of
a typical cloud in the active galactic nuclei (AGN) broad-line
regions (BLRs)upon the incident
flux of ionizing continuum can be nonlinear. We study how this
nonlinearity can be taken into account in estimating the size of
the BLR by means of the ``reverberation'' methods. We show that
the BLR size estimates obtained by cross-correlation of
emission-line and continuum light curves can be much (up to an
order of magnitude) less than the values obtained by reverberation
modelling. This is demonstrated by means of numerical
cross-correlation and reverberation experiments with model
continuum flares and emission-line transfer functions and by means
of practical reverberation modelling of the observed optical
spectral variability of NGC 4151. The time behaviour of NGC 4151
in the H$\alpha$ and H$\beta$ lines is modelled on the basis of
the observational data by Kaspi~{\it et al.} (1996) and the
theoretical BLR model by Shevchenko (1984, 1985). The values of
the BLR parameters are estimated that allow to judge on the size
and physical characteristics of the BLR. The small size of the
BLR, as determined by the cross-correlation method from the data
of Kaspi~{\it et al.} (1996), is shown to be an artifact of this
method. So, the hypothesis that the BLR size varies in time is not
necessitated by the observational data.
\end{abstract}


\section{Introduction}

In the early 1970's, in the course of observations of rapid
variability of the optical spectrum of the Seyfert galaxy
NGC~4151, the time lag of variations in the H$\alpha$ line
with respect to variations in the optical continuum was
discovered \citep{lyutyi71, cherepashchuk73}. The time lag was
interpreted by \citet{lyutyi77, lyutyi82} as a consequence of the
fact that the emission-line clouds are at some distance from the
ionizing radiation source. Later on, \citet{antonucci83} observed
much smaller time lags in variations of the H$\beta$ and H$\gamma$
lines. Such a difference in the time lag, in H$\alpha$ greater
than in H$\beta$, is observed in other active galactic nuclei (AGN) as
well (see Table~6 in \citealt{peterson04}). According to
\citet{shevchenko84, shevchenko85a},  this difference in time
lag is due to an essential nonlinearity in the dependence of the
H$\alpha$ luminosity of an individual cloud upon the ionizing
continuum flux incident on the cloud, the dependence in the
higher order Balmer lines being close to linear. This explanation
was made with the assumption that the duration of the
emission-line flare is much greater than the duration of the flare
in the ionizing continuum, and the duration of the latter one
allows its description by $\delta$-function. However, continuum
variations are not so rapid usually; therefore, in order to
extract physical information from the observed emission-line
variations, it is necessary, in addition to theoretical estimates,
to use numerical modelling taking into account the timescale
of continuum variations.

In the present paper, we study theoretically how the nonlinearity
in the emission-line luminosity, $L_\mathrm{l}$, of the broad-line
region (BLR) cloud, in its dependence on the ionizing continuum flux,
$F_\mathrm{i}$, incident on the cloud, can be taken into account
in estimating the BLR sizes by means of the ``reverberation''
methods. We show that the BLR size estimates obtained by
straightforward cross-correlation of emission-line and continuum
light curves can be much (up to an order of magnitude) less than
those obtained by reverberation modelling. First of all, we
demonstrate this by means of abstract representative numerical
cross-correlation and reverberation experiments with model
continuum flares and emission-line transfer functions. Then we
accomplish practical numerical modelling of the light curves of
NGC~4151 in the H$\alpha$ and H$\beta$ lines on the basis the
observational data by \citet{kaspi96} and the theoretical model of
the BLR by \citet{shevchenko84, shevchenko85a}. This model is
characterized by allowing for thick geometries of the BLR, taking
into account the anisotropy of line emission of individual clouds
and, most important, taking into account the nonlinearity of the
``$L_\mathrm{l}$--$F_\mathrm{i}$'' relation. This nonlinearity allows
one to explain the differences in the time lags for different
lines. Cross-correlation estimates of the BLR size are also made.
They turn out to be small in comparison to the estimates obtained
by the direct reverberation modelling.

The values of the parameters of the BLR model are derived directly
from the reverberation modelling, and that is why we do not use
any specific numeric results of modern photoionization models of
the emission-line spectra of AGN. We use only the fact that,
according to these models, the emission-line response of an
individual cloud (particularly, in H$\alpha$ and H$\beta$) can be
nonlinear. We also allow for a constant emission-line component.

Though the presented theoretical inferences on the time lags can
be of general interest, our primary goal is to apply them to
explaining the time behaviour of the H$\alpha$ and H$\beta$ lines
in the emission-line spectrum of NGC~4151, in the framework of a
simple uniform one-component model. Of course, explaining the time
behaviour of the whole emission-line spectrum
would require much more complicated multi-component models.

\section[]{Effective Stratification of a Homogeneous BLR}

The broad-line region (BLR) of an active galactic nucleus,
according to the ``standard model'', see, e.g.
\citealt{peterson88}, represents an aggregate of line-emitting
clouds under the effect of ionizing radiation of the central
source.

The dependence of the emission-line luminosity $L_\mathrm{l}$ of an
individual cloud upon the value of the incident ionizing flux
$F_\mathrm{i}$, in accordance with the photoionization models of
spectra of active galactic nuclei (see, e.g.,
\citealt{kwan84, mushotzky84}), is described by a power law:
$L_\mathrm{l} \propto F_\mathrm{i}^s$, where $s \ge 0$.

The rate of heat input in a gas cloud optically thick in the
ionizing continuum is directly proportional to the value of the
ionizing flux incident on this cloud. \citet{kwan84} noted that
therefore the cloud's emission-line luminosity should be, in a
first approximation, directly proportional to the ionizing flux;
different lines, however, behave differently. For example, in the
case of L$\alpha$ the dependence is somewhat weaker than linear.
The L$\alpha$ quanta leaving the cloud are produced in the
traditional \mbox{H\,{\sc ii}} zone. At the high ionization
parameters typical of AGNs the collisional ionization from the
excited levels of hydrogen (in particular, from the second level)
are effective even in the \mbox{H\,{\sc ii}} zone. With increasing
ionization parameter their efficiency grows, and this leads to
weakening of the specified dependence \citep{kwan81,kwan84}.

So, the emission-line response of an individual cloud can be
nonlinear. This fact was recognized already in the first
successful photoionization models. According to them, quanta in
many lines are produced mainly not in the traditional
\mbox{H\,{\sc ii}} zone, but deeper, in the so-called ``deep
partly ionized zone''. Successful modelling of stationary optical
emission-line spectra of AGN requires the following two
circumstances to be taken into account \citep{kwan79}: the
power-law shape of the spectrum of ionizing continuum (i.e., the
fact that the major fraction of ionizing quanta is in the X-ray
part of spectrum) and the big column densities of the clouds
emitting in lines. If the X-ray luminosity of the ionizing source
is great enough in comparison with the UV one, the ``deep partly
ionized zone'' is formed in the cloud. Taking into account the
contribution of this zone increases the luminosity of the cloud in
Balmer lines, whereas the luminosity in L$\alpha$ is stabilized at
the level of the luminosity of the \mbox{H\,{\sc ii}} zone. So,
the collisional amplification of Balmer and Paschen lines takes
place in the ``deep partly ionized zone''. Inside this zone the
excitation temperatures of these lines increase with optical
depth, but ultimately attain some limiting values. The limiting
values are insensitive to variation of the ionizing flux, because
the Balmer and Paschen continua dominate in cooling at such depths
\citep{kwan84}. According to \citet{kwan81}, when
collisional ionization becomes the main source of ionizations and
cooling, the rate of cooling increases with increasing electron
temperature approximately as $\exp(-32 \times 10^4\,{\rm
K}/T_\mathrm{e})$. In the standard model by \citet{kwan81},
$T_\mathrm{e} \approx 8000$~K in ``the deep zone''; the steep
dependence of the rate of cooling on temperature, as Kwan and
Krolik noted, provides only weak variation of $T_\mathrm{e}$ with
depth and insensitivity of $T_\mathrm{e}$ to variation of the model
parameters, in particular, the ionization parameter. Increasing
the ionizing flux makes higher levels of hydrogen attain the
limiting excitation temperatures; the luminosity of the ``deep
zone'' in the relevant lines then ceases to react to changes of
the ionizing flux, i.e., in this limit they are constant. In the
Balmer series, the approach to the limiting temperatures affects
first of all the H$\alpha$ line, then H$\beta$, and so on. Thus,
according to the photoionization models \citep{kwan81,kwan84}, the
dependence of the cloud's emission-line luminosity on the incident
ionizing flux for the H$\alpha$ line is weaker than for H$\beta$,
for H$\beta$ is weaker than for H$\gamma$, and so on.

Due to the difference between H$\alpha$ and H$\beta$ in the value
of the $s$ parameter, the Balmer decrement increases with increasing
distance of clouds away from the central source,
therefore a photographic BLR image (if such an image could be
obtained) would be larger in H$\alpha$ than in H$\beta$.
A formula for an effective BLR radius in a line with
an arbitrary $s$ value in the homogeneous model of the cloud
aggregate was deduced in \citep{shevchenko85a}. This effective
stratification is explained by differences between emission lines
in the degree of nonlinearity of the $L_\mathrm{l}(F_\mathrm{i})$
function. \citet{shevchenko88} showed that within the framework of
the homogeneous model of the cloud aggregate, if one takes into
account the results of the photoionization calculations of the
emission-line spectra of active galactic nuclei
\citep{kwan84, mushotzky84}, it is possible to explain the
observed time lags and amplitudes of variations in major optical
and ultraviolet (UV) emission lines in the spectrum of NGC~4151.

After the first successes of the photoionization computations of
the AGN emission-line spectra, significant progress was made in
this field; see, e.g., reviews by \citet{ferland03} and
\citet{leighly07}. Multi-component models were proposed and
studied \citep{collin-souffrin88, collin-souffrin88a, korista97},
which allowed to reproduce the relative fluxes in high-ionization
and low-ionization lines simultaneously. Evidence was found for
the presence of optically thin line-emitting gas \citep{ferland90,
shields95}.
This progress promoted much deeper understanding of the AGN
emission-line spectra~--- it turned out that the uniform models
are too simple to reproduce the whole spectra. However, in what
follows, our study concerns only Balmer lines. We aim to explain
the time behaviour of the H$\alpha$ and H$\beta$ lines in the
emission-line spectrum of NGC~4151, in the framework of a simple
uniform one-component model. Of course, explaining the behaviour
of the emission-line spectrum in total may require much more
complicated multi-component models.

Effective, not physical, stratification is present in our
one-component model, due to the nonlinearity in each cloud's line
emission. The alternative to a homogeneous BLR with effective
stratification is a physically stratified BLR.
Investigating variability of the UV lines of NGC~4151,
\citet{ulrich84} offered the BLR model consisting of three zones
with different physical characteristics (see Table~2 in their
article). \citet{gaskell86} proposed a model consisting of two
zones (see Table~1 in their article). These models are not
considered henceforth; we adopt the effective stratification
picture as implied by the nonlinearity in cloud's line emission.

\section[]{The Reverberation Model}

\citet{blandford82} offered a procedure to recover the
BLR structure by analysis of line and continuum lightcurves.
This is the so-called method of
``reverberation mapping''. Its essence consists in the following:
the observed light curve in a line is supposed to represent a
convolution of two curves: the transfer function describing
physical characteristics and the geometry of the BLR and the light
curve in ionizing continuum. The emission-line luminosity of an
individual cloud was supposed to depend linearly on the
incident ionizing flux.

\citet{shevchenko84, shevchenko85a} found necessary and sufficient
conditions for the existence of a time lag of a maximum of
an emission-line flare in relation to a (short duration)
continuum flare when the BLR structure is isotropic with respect
to the central source; these conditions are: the typical cloud
should emit in the line mainly from the side facing the central
source, and, either a central cavity should be effectively present
in BLR, or the $s$ parameter in the formula $L_\mathrm{l} \propto
F_\mathrm{i}^s$ should be less than one. These conditions set useful
reference points for our modelling. In the 1990's the effect of
nonlinear response as well as of anisotropy of the individual
cloud emission in specific BLR models, as applied to the
cross-correlation analysis, was studied in detail by
\citet{sparke93} and \cite{obrien94, obrien95}.

We adopt the homogeneous model of the cloud aggregate
\citep{shevchenko84, shevchenko85a}. The effective BLR radius $R$
is defined by screening of the peripheral part of the aggregate by
the clouds situated closer to its centre: $R = (\sigma n)^{-1}$,
where $\sigma$~[cm$^2$] is the mean geometrical cloud section
orthogonal to direction to the central source, $n$~[cm$^{-3}$] is
the cloud concentration (number of clouds in a unit volume).
Generally, the BLR can contain a central cloud-free cavity of
radius $R_0$. Let us remark that the Balmer quanta, unlike the
ionizing quanta, can leave the BLR freely even at large
cloud-covering factors of the ``sky'' of the central source,
because the dispersion of the cloud velocities in the BLR is
assumed to be great; the latter fact is testified by the large
width of the observed emission lines.

The model of a homogeneous (outside the central cavity)
distribution of the clouds is equivalent, in what concerns the
transfer function form, to a model with zero covering factor but
with $n \propto e^{-r/R}$, an exponential decrease in the
cloud concentration with increasing distance from the centre.
In both interpretations, $R$ characterizes the BLR radius
for all lines. In the first case, it is the radius of the ``lit''
zone in the homogeneous aggregate, and in the second case it is
the $e$-folding scale of the cloud concentration.

We assume that a typical BLR cloud represents a flat ``pancake''
emitting lines solely from the side facing the ionizing radiation
source, and, what is more, emitting orthotropically. See
discussion in \citep{shevchenko85b} on the physical basis for this
assumption. The planes of the clouds are either orthogonal to the
direction to the central source, or are oriented randomly. The
phase function, describing the phase angle dependence of the
cloud's line emission, is different in these two cases. For
random orientation, the effective phase function (the phase
function of a volume unit containing many clouds) coincides with
the phase function of a spherical cloud, provided
\citep{shevchenko85b}: the cloud is completely opaque in the line,
the line quanta are produced at small optical depths, and the
cloud surface emits in the line orthotropically.

One should make a reservation that the pancake shaped cloud, as
well as a uniform cloud aggregate model itself, is a physical
idealization that can be used only as an approximation for the
real arrangement of line-emitting material in the BLR. The real
structure might be closer to a combination of a disk and an
outflowing wind \citep{emmering92, murray95, chiang96, bottorff97,
elvis00}. We adopt the pancake shape for the BLR cloud exclusively
for convenience in mathematical modelling: indeed, according
to \citet{shevchenko85b}, the phase function of the pancake cloud
with the plane orthogonal to the ionizing source direction
provides the maximum anisotropy of line emission, i.e., this is a
physical limit worth theoretical examination, while the phase
function of a spherical cloud (or, equivalently, randomly oriented
pancakes) gives an approximation for the phase function of
randomly oriented optically thick line-emitting material, i.e., it
describes a situation that is expected to be closer to reality.

If the ``pancakes'' are orthogonal to the central source
direction, the transfer function representing the dependence of
the observed integrated emission-line flux $f(t)$ on time $t$
counted from the moment of the $\delta(t)$--flare of the central
source in continuum, is as follows \citep{shevchenko84,
shevchenko85a}:

\begin{equation}
\label{transfer1}
f(t) \propto \left\{
\begin{array}{cl}
0, & 0 \le t \le {\displaystyle R_0}, \\
&\\
{\displaystyle R^{-1} \int\limits_{R_0}^{t}} g(r,t)\,dr,
& {\displaystyle R_0 \le t \le 2 R_0},\\
&\\
{\displaystyle R^{-1} \int\limits_{t/2}^{t}} g(r,t)\,dr, & t \ge
{\displaystyle 2 R_0},
\end{array}
\right.
\end{equation}

\noindent
where
\[
g(r,t) = \left(\frac{t}{r} -1 \right) r^{1-2s} e^{-r/R},
\]

\noindent and $r$, $R$, $R_0$ are measured in the light-travel
time units.

In the case when the planes of clouds are oriented randomly,
their mean phase function coincides with the phase function of a
spherical cloud. This function is as follows
\citep{shevchenko85b}:

\begin{equation}
\label{pf2} j(\theta) \propto (1 + \cos \theta) \left(1 +
\frac{s}{2}\cos \theta \right),
\end{equation}

\noindent
where $\theta$ is the ``ionizing source -- cloud --
observer'' angle, $0 \le \theta \le \pi$, $0 \le s \le 2$. General
formula~(3) in \citep{shevchenko84} for the transfer function,
after substitution of phase function~(\ref{pf2}), becomes:

\begin{equation}
\label{transfer2} f(t) \propto \left\{
\begin{array}{cl}
{\displaystyle\frac{t}{R}\int\limits_{R_0}^\infty} g(r,t)\,dr,
& 0 \le t \le {\displaystyle 2 R_0},\\
&\\
{\displaystyle\frac{t}{R}\int\limits_{t/2}^\infty} g(r,t)\,dr, & t
\ge {\displaystyle 2 R_0},
\end{array}
\right.
\end{equation}

\noindent
where
\[
g(r,t) = \left(1 + {\displaystyle\frac{s}{2}\left(\frac{t}{r} -
1\right)}\right) r^{-2s} e^{-r/R}\>.
\]

\begin{figure}
\begin{center}
\includegraphics[width=130mm]{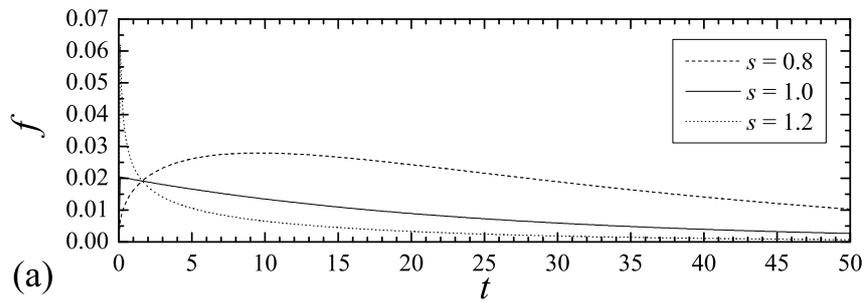}\\
\includegraphics[width=130mm]{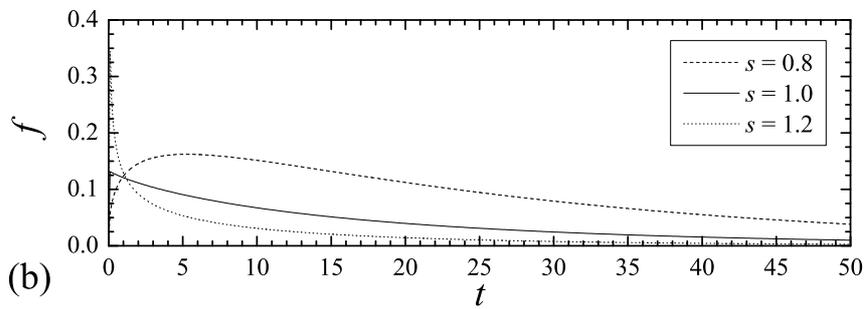}\\
\caption{Transfer functions~(\protect\ref{transfer1}) and
(\protect\ref{transfer2}) (figures (a) and (b), respectively) for
three values of the $s$ parameter, while $R_0=0$, $R=15$ lt-days.}
\label{fig1}
\end{center}
\end{figure}

\noindent Transfer functions~(\ref{transfer1}) and
(\ref{transfer2}) can be expressed through incomplete
$\gamma$ functions. The behaviour of the transfer functions with
different values of the $s$ parameter (while $R_0=0$, $R=15$
lt-days) is demonstrated in Fig.~\ref{fig1}. The qualitative
difference in the behaviour of the functions with $s$ less and
greater than unity is clearly seen. In particular, $f(t)$ peaks at
$t = 0$ for $s \ge 1$ and at $t>0$ for $s<1$.

The model emission-line light curve is determined by the
convolution formula:

\begin{equation}
F_\mathrm{l}(t) = a \int_0^\infty f(\tau) F_\mathrm{c}^s(t - \tau)d\tau \>,
\label{conv}
\end{equation}

\noindent where $F_\mathrm{l}$ is the integrated flux in the line,
$F_\mathrm{c}$ is the observed flux in continuum, $a$ is the
normalizing dimensional factor. Since it is the optical continuum,
not the ionizing one, that is observed, we use an assumption,
formulated below, on a relation between the continua. A difference
of expression~(\ref{conv}) from those usually used (valid in the
case of the linear ``$L_\mathrm{l}$--$F_\mathrm{i}$'' relation; see,
e.g., \citealt{blandford82} and \citealt{horn04}) consists in
raising of $F_\mathrm{c}$ to the power $s$. Let us remark that,
according to~(\ref{conv}), on taking $f(\tau)$ in the form of
$\delta$-function, one gets $F_\mathrm{l}(t) \propto F_\mathrm{c}^s(t)$,
i.e., the dependence for the case of quasi-stationary spectrum;
see \citep{shevchenko88}.

\section[]{The time lag and the cross-correlation method}
\label{tlccm}

Techniques for cross-correlation analysis of AGN
emission-line variability have demonstrated remarkable progress
during the last decade. The methods of calculation of the basic
properties of the cross-correlation function (CCF), namely, the lags of
the CCF peak and CCF centroid and their uncertainties, were
greatly improved (e.g., \citealt{white94, peterson95, peterson98,
welsh99}). In particular, it was realized that the CCF peaks and
centroids underestimate the BLR size \citep{perez92, welsh99}, and
that taking into account the continuum variability time scale is
important for correct estimation of the BLR size (e.g.,
\citealt{edelson88}).

Let us consider the time lag as determined by means of
cross-correlation analysis in the case of nonlinear emission-line
response of an individual cloud. In this Section, we measure the
time lag in model numerical experiments and study the dependence
of the time lag on the parameters of a model transfer function and
duration of the continuum flares. We consider the case of a single
flare of various durations. As the model transfer function we take
Eq.~(\ref{transfer1}) corresponding to the case of the ``pancake''
clouds orthogonal to the central source direction. The central
cavity in the BLR is set to be absent: $R_0 = 0$.

The model light curve in the continuum is assumed to have the form
of the bell-like function $F_\mathrm{c}(t) = \mbox{sech} ((t -
t_0)/T)$, where $t_0 = 50$~d and $T$ is effective duration of the
flare, $t$ is time in days. The model emission-line light curves
are computed on the time interval of 500~d with the step of
$0.05$~d.

\begin{figure}
\begin{center}
\includegraphics[width=130mm]{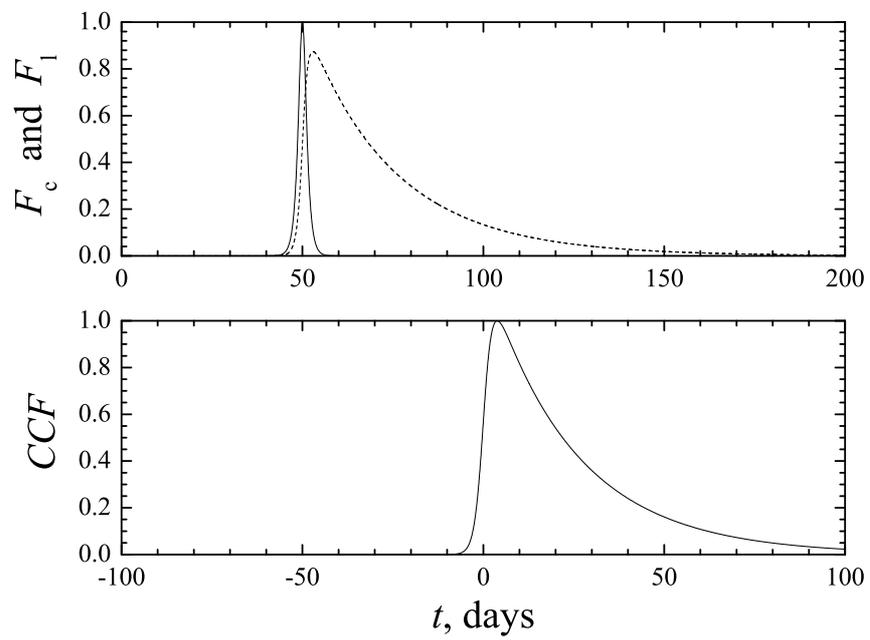}\\
\caption {The upper plot: the model light curve in the continuum
(the continuous line; $R=15$ lt-days, $R_0=0$, $T=1$~d, $s=1$) and
the computed emission-line light curve (the dashed line). The
lower plot: the normalized cross-correlation function.}
\label{fig2}
\end{center}
\end{figure}

In the upper part of Fig.~\ref{fig2}, the model light curve in the
continuum ($T = 1$~d) and the computed emission-line light curve
are presented. The latter curve has been obtained by means of
convolution of the light curve in the continuum and the transfer
function~(\ref{transfer1}) with $R = 15$~lt-days, $R_0=0$, $s =
1$. In the lower part of Fig.~\ref{fig2}, the normalized
cross-correlation function of these curves is plotted. The shift
of the peak of the cross-correlation function is clearly visible;
as determined numerically, $\Delta t_\mathrm{peak} \simeq 3.8$~d.
Note that the exponential-like decay of the resulting curves in
both plots reflects the radial structure of the line-emitting
region, and the rise in the CCF reflects the shape of the
continuum flare.

One may argue that the linear response model is just a
linearization of the nonlinear response model, and that any
BLR radius estimate made in the linear response model is
therefore an approximation that might be not far from reality.
However, one should take into account, firstly, that the ``equal
time-travel'' paraboloidal surface inside the BLR covers a whole
range of distances from the ionizing source just after the
ionizing flare, secondly, that with increasing time after the
flare this surface retreats from the source to larger
distances. The slopes of the linearized dependences for individual
clouds on the surface vary significantly in both space and time.
The response slopes might be averaged on the surface, but then the
change of the averaged slope with time should be taken into
account; the latter is never done in practice. So, it is not
surprising that the nonlinear response model, as compared to the
linear one, can give very different quantitative results on the
BLR radius. This directly follows from the qualitative differences
in the response function for different values of the $s$
parameter, as seen in Fig.~\ref{fig1}. A vivid manifestation of
the insufficiency of the linearized response model is that
increasing the time lag value in the nonlinear response model can
be achieved either by increasing the BLR radius or by specific
increasing the response nonlinearity, namely, by decreasing the
$s$ parameter value in relation to unity (see
relation~(\ref{deltat})), while in the linear response model only
the BLR radius can be varied.

\citet{shevchenko85a, shevchenko94} obtained an approximate
theoretical relation of the time lag of the maximum of the
emission-line light curve to the $s$ parameter in the homogeneous
model of the cloud aggregate with or without a central cavity
($R_0 \ge 0$). The continuum flare was described by a
$\delta$ function. This relation is as follows:

\begin{equation}
\Delta t = \left\{ \begin{array}{ll}
W (1-s) R, & {0 \le s \le {1 - 2 R_{0}/(W R)},} \\
2R_{0},          & {s \ge {1 - 2 R_{0}/(W R)},}
\end{array}
\right.
\label{deltat}
\end{equation}

\noindent where the constant $W$ depends on the choice of phase
function; $W = 3.19$ in the considered case of clouds with regular
orientation. In the case of phase function~(\ref{pf2}) one has $W
= 2$.

\begin{figure}
\begin{center}
\includegraphics[width=130mm]{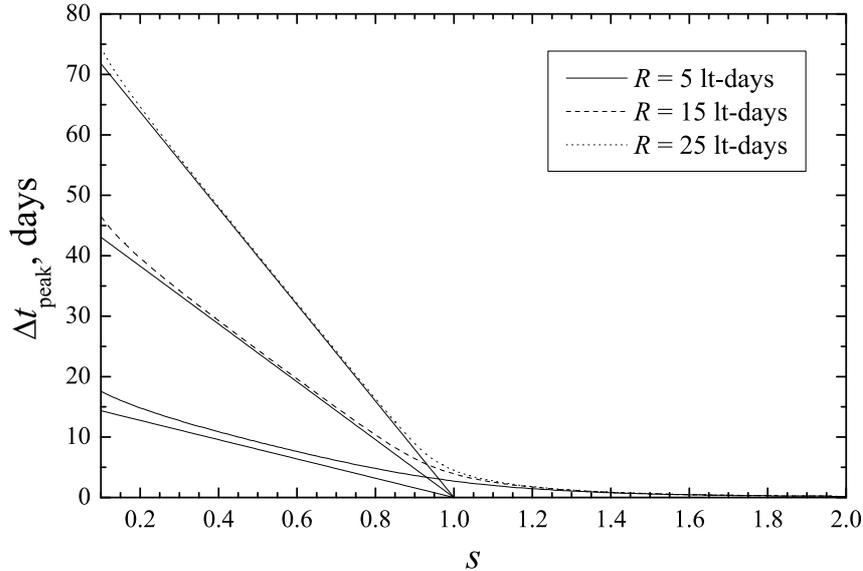} \\
\caption{The dependence of the time lag on the parameter $s$;
$R_0=0$, $T=1$~d. The straight line segments represent theoretical
relations~(\protect\ref{deltat}).}
\label{fig3}
\end{center}
\end{figure}

By examining the shifts of peaks of cross-correlation
functions at various values of parameters we can assess how well
relation~(\ref{deltat}) works when the ionizing flare
has a finite duration. Consider first how
the time lag varies with $s$ at fixed $R$ (Fig.~\ref{fig3}).
The time lag, as defined here, is the value
of the distance along the time axis from $t = 0$ up to the first
maximum of the cross-correlation function; i.e., it is $\Delta
t_\mathrm{peak}$. We do not examine the shift of the centroid of the
cross-correlation function here. The parameter $s$ is varied from
$0.1$ to $2.0$ with the step of $0.01$. The ionizing flare
duration is fixed at $T=1$~d. The curves for $R = 5$, 15 and
25~lt-days are plotted. Theoretical dependences~(\ref{deltat}) for
$R = 5$, 15 and 25~lt-days and $R_0 = 0$ are plotted as straight
line segments. It is clear that the theoretical and numerical
estimates of the time lag at such a relatively small duration of
the continuum flare are in a good agreement.

\begin{figure}
\begin{center}
\includegraphics[width=130mm]{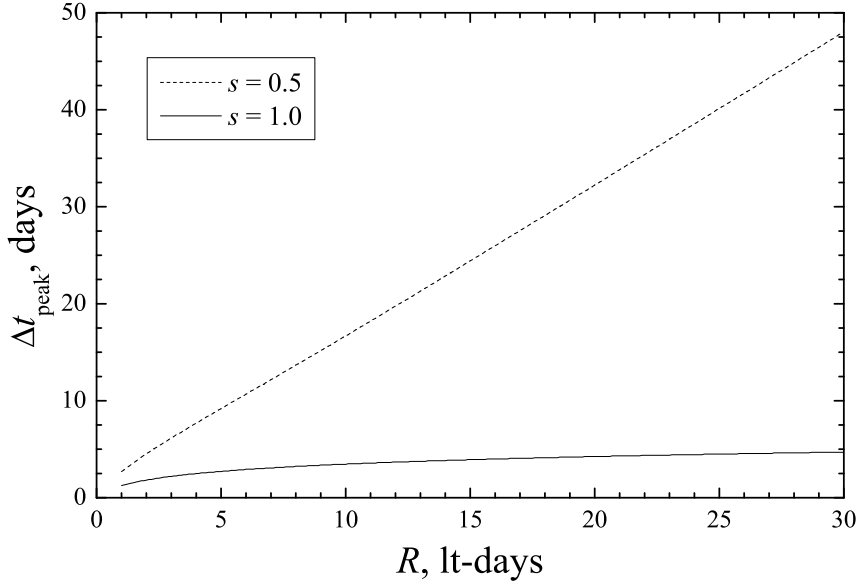} \\
\caption{The dependence of the time lag on the BLR radius $R$;
$R_0=0$, $T=1$~d.}
\label{fig4}
\end{center}
\end{figure}

Now we consider the dependence of the time lag on the BLR radius
$R$ (Fig.~\ref{fig4}). The ionizing flare duration is the same,
$T=1$~d. The dependences for $s=0.5$ and $s=1$ are plotted. We see
that the linear character of theoretical relation~(\ref{deltat}),
valid for the $\delta(t)$ ionizing flare, is preserved in the case
of $s=0.5$. In what concerns the $s=1$ case, it is completely
different. From the viewpoint of relation~(\ref{deltat}), this
case is degenerate, and the predicted value of $\Delta
t_\mathrm{peak}$ is constant (zero). In reality we observe the
``$\Delta t_\mathrm{peak}$--$R$'' relation similar to a logarithmic
one.

So, as follows from Fig.~\ref{fig3} and Fig.~\ref{fig4}, for $s
\ge 1$ the CCF $\Delta t_\mathrm{peak}$ value depends only
weakly on $R$. The cross-correlation peak time lag for $s \ge 1$
is small in comparison with $R$ expressed in the light-travel time
units. The difference can reach an order of magnitude.
For $s = 1$, a similar phenomenon was observed by \citet{perez92} for
the CCF centroid estimates of the BLR size in an isotropic (with
respect to the ionizing source) model of the line-emitting cloud
distribution. They found that such estimates can be less than the
real size of the BLR, the difference reaching two times. This is
in accord with our findings.

\begin{figure}
\begin{center}
\includegraphics[width=130mm]{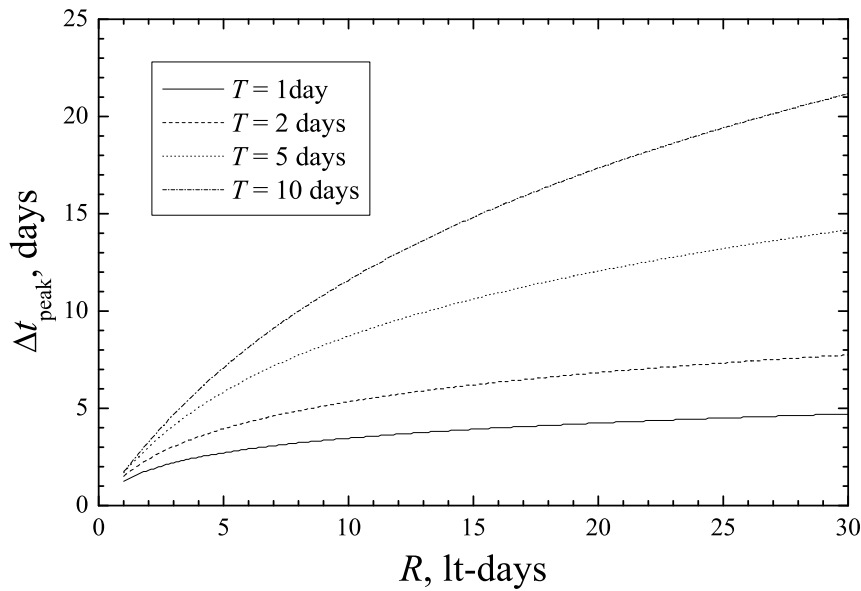} \\
\caption{The dependence of the time lag on the BLR radius $R$ at
various values of duration $T$ of the ionizing flare; $s = 1$,
$R_0=0$.}
\label{fig5}
\end{center}
\end{figure}

What is the role of the ionizing flare duration in the degenerate
case $s = 1$? The dependences of the time lag on $R$ in this case
for various fixed values of $T$ are plotted in Fig.~\ref{fig5}. We
see that this role is far greater than that of the BLR radius. The
observed ``$\Delta t_\mathrm{peak}$--$R$'' dependences seem to be
logarithmic indeed; one can verify that, unlike the rational and
power-law functions, the functions of the form $a + b \ln (R +
c)$, where $a$, $b$, $c$ are fitting parameters, provide ideal
visual description of the observed curves. Note that further
work is required to explore how the logarithmic dependence found
here is vulnerable to the choice of the ionizing flare shape.

In summary, according to our numerical-experimental findings, the
BLR radius is only weakly related to the measured $\Delta
t_\mathrm{peak}$ value in the mathematically degenerate but
observationally most common case of $s=1$. The role of the
timescale of variability is far greater. Therefore, the lines with
$s \approx 1$ are of little help in determining the BLR size by
means of cross-correlation techniques; instead, the lines with $s$
essentially less than 1, such as H$\alpha$, should be used for
this purpose. This conclusion has been obtained in a model
framework and thus may be model-dependent in some way. However, it
makes clear that there are no general theoretical grounds to
believe that $\Delta t_\mathrm{peak}$ is mostly determined by the BLR
size. Note that solely the CCF peak offset has been examined.
The CCF centroid offset should be examined as well in a future
study to check whether it exhibits the same behaviour.

\section[]{Reverberation modelling of the emission-line light curves of NGC~4151}

In this Section we examine the effect of taking into account the
nonlinearity in a cloud's line emission in practical modelling of
emission-line variability of an AGN. We model the
emission-line variability of the Seyfert galaxy NGC~4151. The
observational data of \citet{kaspi96} on variability of the
nucleus of this galaxy in H$\alpha$, H$\beta$ and optical
continuum is used. The observations of \citet{kaspi96}, performed
in the framework the AGN Watch Programme, cover the time interval
of approximately three months, from 1993 November until 1994
February. While we use the light curve continuum data for the
whole time span of the observations, the data on the fluxes in the
lines during the first five days and during the last five days of
the observational time span are excluded, following the usual
practice of eliminating the border effects (see~\citet{maoz91}).
The error bars of the individual observations, defining the
weights of the observations, are taken into account in the
modelling.

We optimize the model parameters by using a non-linear
least-squares method \citep{levenberg44,marquardt63} to minimize
$\chi^2$, thereby finding best-fit parameter values and their
standard errors (see~\citet{press97}). We find that the
iterations converge to the same solution for random starting
values in ranges specified below, with the deepest minimum of
$\chi^2$ in all cases, though several other local minima exist, as
demonstrated below. We expand the standard errors by the
square root of the reduced $\chi^2$, because the best-fit reduced
$\chi^2$ is greater than unity in all best-fit models found.

As the transfer functions, we use Eqs.~(\ref{transfer1}) and
(\ref{transfer2}). The best approximation of the computed model
light curve to the observed one is found by means of the
modelling. We vary six parameters: the BLR radius $R$,
the nonlinearity parameters
$s_\mathrm{opt}$ for H$\alpha$ and H$\beta$, the normalizing
coefficients $a$ for H$\alpha$ and H$\beta$, the flux $F_{\alpha
\mathrm{n}}$ in the narrow component of H$\alpha$ (on the $F_{\beta
\mathrm{n}}$ value for H$\beta$ see below).

The radius $R_0$ of the central cavity (a zone free from
line-emitting clouds) has not been varied because it is already
known to be most probably small. A comparison of the known values
of time lags in different Balmer lines of NGC~4151 allowed
\citet{shevchenko85a} to conclude that the upper bound of the
cavity radius $R_0$ is 4--5 times less than the effective BLR
radius. The deduction that the central cavity is small is in
agreement with conclusions by \citet{maoz91} and \citet{xue98}.
Note that the radius of the accretion disc in the centre of the
nuclear region is estimated to be equal to $0.6$--2~lt-days
\citep{lyuty05, sergeev05, sergeev06}. Taking into account the
uncertainty of this estimate and the uncertainty of the $R_0$
estimates, the probable existence of this component of the nuclear
region does not at all contradict the conclusions on the small
relative size of the central cavity in the BLR.

For the continuum flux we take the flux at the wavelengths of
4560--4640\AA \ (the ``4600\AA'' region), because this region
corresponds to the shortest wavelengths at which \citet{kaspi96}
measured the continuum flux. The series of the observed values
of the continuum flux are recalculated for presentation on the
uniform time grid by means of cubic spline interpolation.

The constant contribution to the integrated flux in H$\alpha$ and
H$\beta$ due to the narrow components of the lines has been taken
into account in the following way. We set $F_{\alpha}(t) =
F_\mathrm{c}^{s_\alpha}(t) + F_{\alpha \mathrm{n}}$ and $F_{\beta}(t) =
F_\mathrm{c}^{s_\beta}(t) + F_{\beta \mathrm{n}}$. The contributions of
the narrow components $F_{\alpha \mathrm{n}}$ and $F_{\beta \mathrm{n}}$
are connected to each other via the constant Balmer decrement
$D_\mathrm{n} = F_{\alpha \mathrm{n}}/F_{\beta \mathrm{n}}$. Therefore in
the course of searching for the best model it is enough to vary
the value of $F_{\alpha \mathrm{n}}$; the value of $F_{\beta
\mathrm{n}}$ is determined via the Balmer decrement. For each of the
considered cases of orientation of the planes of clouds we have
accomplished the modelling twice, namely, for the two reported
values of the Balmer decrement in the narrow components:
$D_\mathrm{n} = 4.47$ as given in Table~1 in \citep{ferland82} and
$D_\mathrm{n} = 7.55$ as given in Table~1 in \citep{sergeev01}. The
difference in the observed values of the decrement $D_\mathrm{n}$ may
reflect either the difficulty in its evaluation or its probable
long-term variability.

The contribution of the stellar component to the observed
continuum flux has been taken into account by its subtraction from
the observed flux prior to modelling. According to
\citet{peterson95} and \citet{kaspi96}, the stellar component
contribution at the wavelength of 4600\AA \ and the aperture used
at their observations is approximately equal to $2.2 \times
10^{-14} \mbox{ erg cm$^{-2}$ s$^{-1}$ \AA$^{-1}$}$.

As it is known from observations (see, e.g., \citealt{crenshaw96,
peterson02}), the light curves of an active galactic nucleus in
the optical and UV continua can be rather different, and the
relation between the continua is nonlinear. For the Seyfert galaxy
NGC~5548, most studied in this respect, the slow components of
variability in the optical and UV continua are connected by the
power law $F_{\mathrm{opt}} \propto F_{\mathrm{UV}}^\gamma$, where
$\gamma \approx {0.56}$ \citep{peterson02}.
Basing on this relation, we find the real values of the $s$
parameter from the values obtained in our modelling of the optical
light curves (we designate these values by $s_{\mathrm{opt}}$)
by means of the formula $s = \gamma s_{\mathrm{opt}}$, where
we set $\gamma = 0.6$.

The value of $R$ does not depend on the line choice. The
values of $s_{\mathrm{opt}}$ for different lines are generally
different, and the same is true for $a$. The initial data for the
iterations of the Levenberg--Marquardt algorithm have been taken
randomly in the following limits: $R$~--- from 1 to 30~lt-days;
$s_{\mathrm{opt}}$ in the both lines~--- from $0.2$ to $2.0$; $a$~---
from $0.01$ to $2.0$; $F_{\alpha \mathrm{n}}$~--- from 0 to $30
\times 10^{-12} \mbox{ erg cm$^{-2}$ s$^{-1}$}$.

\begin{figure}
\begin{center}
\includegraphics[width=130mm]{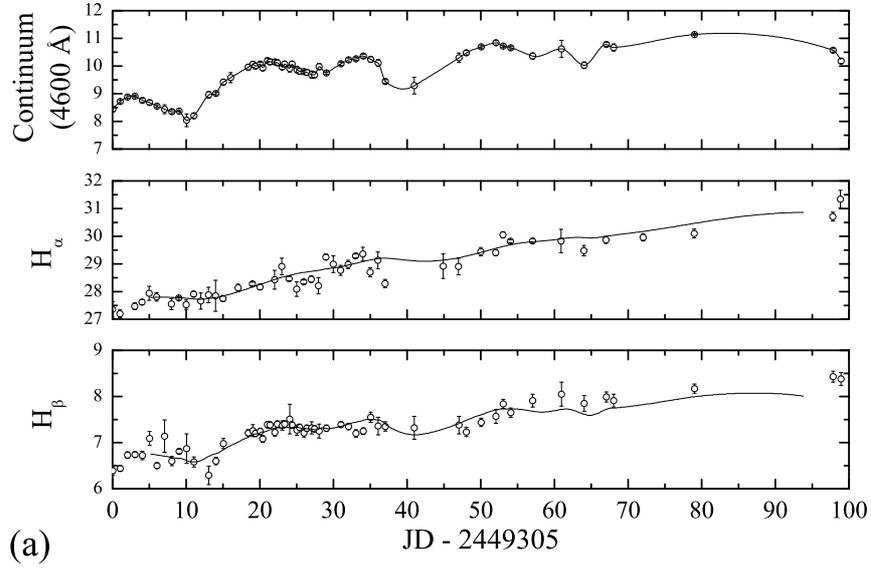}\\
\includegraphics[width=130mm]{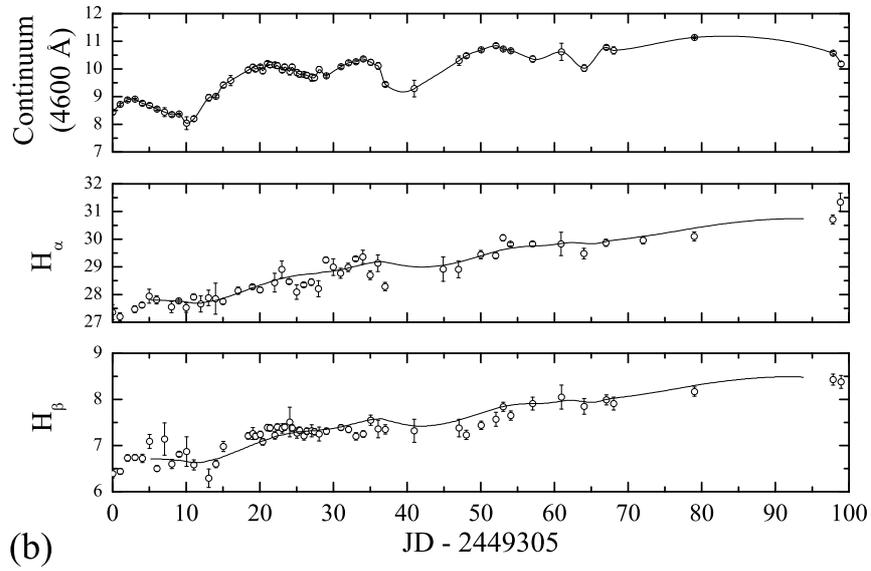}\\
\caption{The best model light curves of NGC~4151 in H$\alpha$ and
H$\beta$ in the case when the planes of clouds are orthogonal to
the direction to the central source. $D_\mathrm{n} = 4.47$ (a),
$D_\mathrm{n} = 7.55$ (b). The continuum flux is given in the
units of $10^{-14} \mbox{ erg cm$^{-2}$ s$^{-1}$ \AA$^{-1}$}$. The
emission-line fluxes are given in the units of $10^{-12} \mbox{
erg cm$^{-2}$ s$^{-1}$}$.}
\label{fig6}
\end{center}
\end{figure}

In the case when the planes of clouds are orthogonal to the
direction to the central source, the best-fit model light curves
are presented in Fig.~\ref{fig6} for two different values of
$D_\mathrm{n}$. The circles designate the observed values of the flux
in optical continuum and the observed integrated emission-line
fluxes according to the data in (\citealt{kaspi96}, Table~2). In
both parts of the Figure, the continuous curve in the upper plot
is the spline interpolation of the continuum flux. The flux is
given in the units of $10^{-14} \mbox{ erg cm$^{-2}$ s$^{-1}$
\AA$^{-1}$}$. In the lower two plots, the best model light curves
in H$\alpha$ and H$\beta$ are presented as continuous curves; here
the integrated emission-line flux is given in the units of
$10^{-12} \mbox{ erg cm$^{-2}$ s$^{-1}$}$.

\begin{figure}
\begin{center}
\includegraphics[width=130mm]{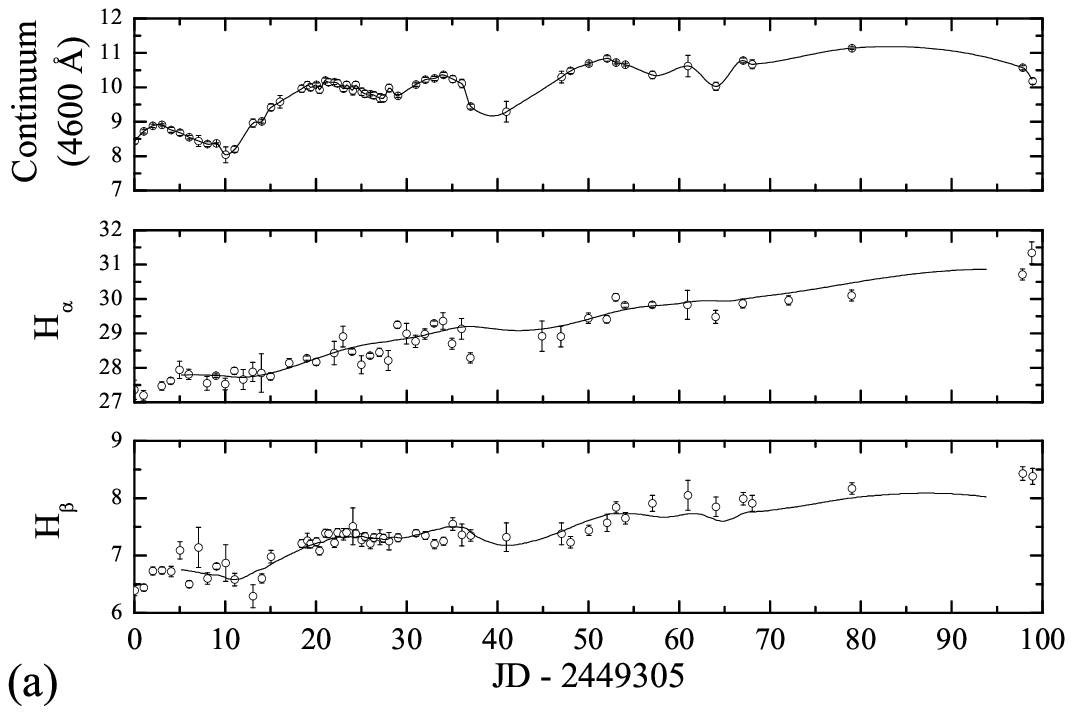}\\
\includegraphics[width=130mm]{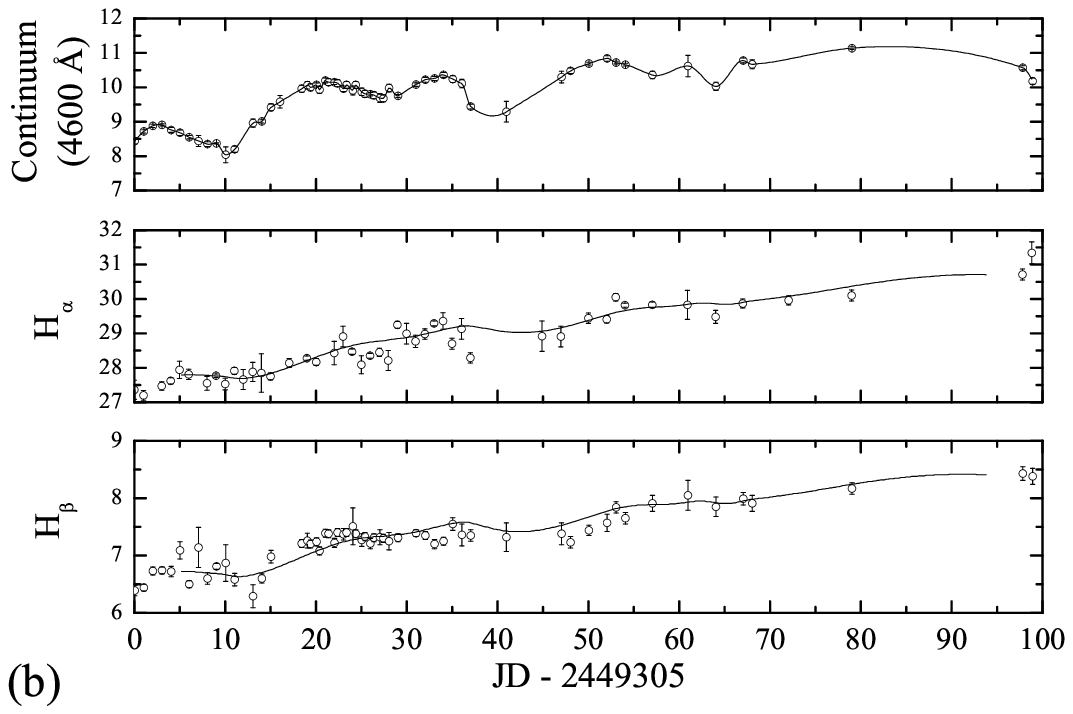}\\
\caption{The same as in Fig.~\protect\ref{fig6}, but the planes of
clouds are oriented randomly.}
\label{fig7}
\end{center}
\end{figure}

In Fig.~\ref{fig7}, the best model emission-line light curves are
presented for the case when the planes of clouds are oriented
randomly.

\begin{figure}
\begin{center}
\includegraphics[width=130mm]{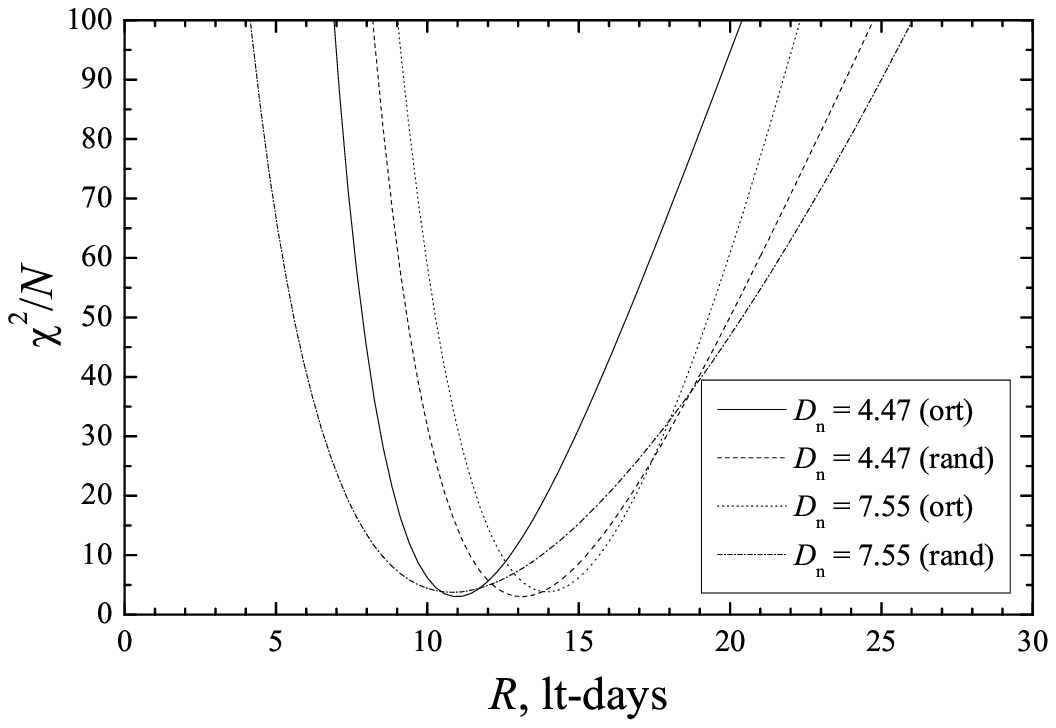} \\
\caption{The dependence of the reduced $\chi^2$ on the BLR
radius $R$ with all other parameters assigned to their best-fit
values.}
\label{fig8}
\end{center}
\end{figure}

\begin{figure}
\begin{center}
\includegraphics[width=100mm]{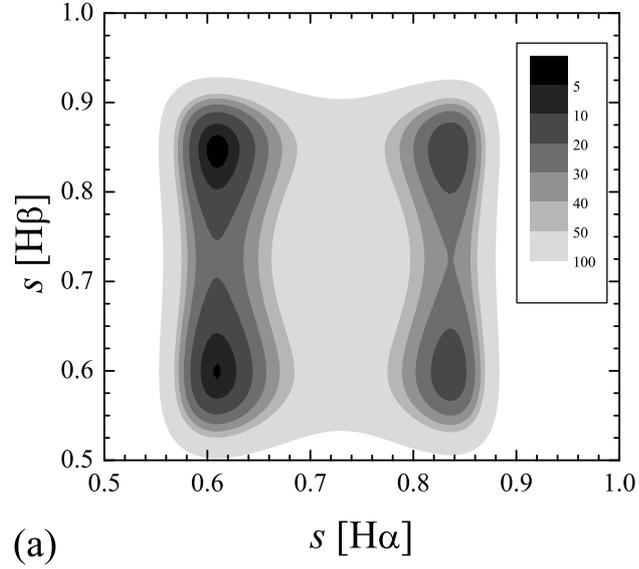}\\
\includegraphics[width=100mm]{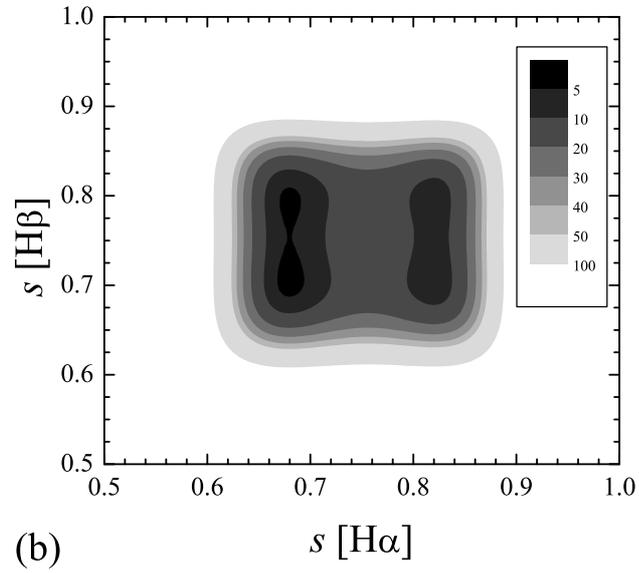}\\
\caption{The dependence of the reduced $\chi^2$ on the
values of the $s$ parameters for H$\alpha$ and H$\beta$ with all
other parameters assigned to their best-fit values. The case when
the planes of clouds are orthogonal to the direction to the
central source. $D_\mathrm{n} = 4.47$ (a), $D_\mathrm{n} = 7.55$ (b).}
\label{fig9}
\end{center}
\end{figure}

\begin{figure}
\begin{center}
\includegraphics[width=100mm]{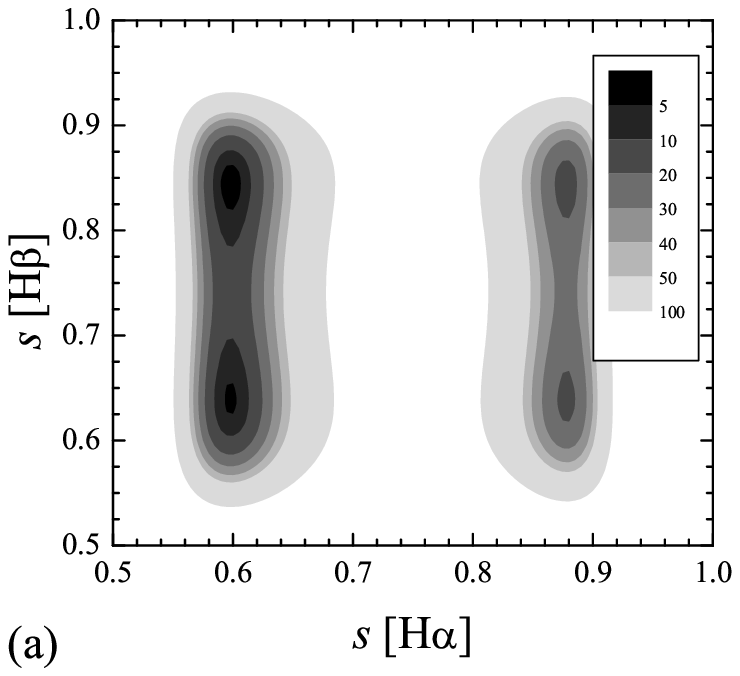}\\
\includegraphics[width=100mm]{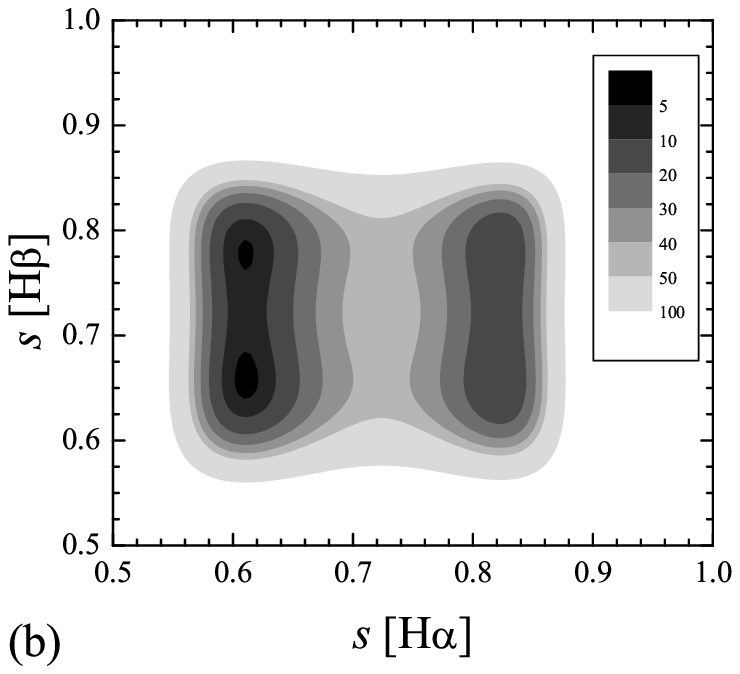}\\
\caption{The same as in Fig.~\protect\ref{fig9}, but the planes of
clouds are oriented randomly.}
\label{fig10}
\end{center}
\end{figure}

To demonstrate the effect of variation of different
parameters in determining the best-fit models, $\chi^2/N$
(where $N$ is the number of degrees of freedom) is shown in
Figs.~\ref{fig8}, \ref{fig9} and \ref{fig10} in dependence on the
main parameters. Fig.~\ref{fig8} gives the dependence of the
reduced $\chi^2$ on the BLR radius $R$ with all other parameters
assigned to their best-fit values. Figs.~\ref{fig9} and
\ref{fig10} show the reduced $\chi^2$ in dependence on the
$s$ parameters for H$\alpha$ and H$\beta$ with all other
parameters assigned to their best-fit values. (Note that these
graphs are presented here for illustrative purposes solely; for
quantitative analysis, when offsetting one parameter, one should
re-optimize all the other parameters while holding the one
parameter fixed at the offset value.) In the presented graphs one
can see that the $\chi^2$ dependence on $R$ is characterized by a
single well-defined minimum. The dependences of $\chi^2$ on the
$s$ parameters have four minima each. The Levenberg--Marquardt
algorithm finds the deepest one.

In Tables~\ref{tab1} and \ref{tab2}, the obtained values of the
model parameters, corresponding to the model emission-line light
curves in Fig.~\ref{fig6} and Fig.~\ref{fig7}, are presented
together with the reduced $\chi^2$ values of the models. In total,
the results of our modelling give the following values of the
radius of the BLR of NGC~4151: $R = 11$--14~lt-days, with the
uncertainty of 5--8~lt-days. At $D_\mathrm{n} = 4.47$, the best-fit
values of the $s$ parameter are: $s \approx 0.6$ for H$\alpha$
and $s \approx 0.85$ for H$\beta$. From comparison of
Tables~\ref{tab1} and \ref{tab2} one can see that at increasing
decrement $D_\mathrm{n}$ from $4.47$ up to $7.55$ the difference in
the computed values of the $s$ parameter for H$\alpha$ and
H$\beta$ becomes less, though does not seem to disappear
completely.

\begin{table}
\begin{center}
\caption{The recovered values of the BLR parameters in the case of
$D_\mathrm{n} = 4.47$} \label{tab1}
\begin{tabular}{@{}lccc}
\hline
\multicolumn{2}{c}{} & Orthog.\ orient.\ & Random orient.\ \\
 && (Fig.~\ref{fig6}a) & (Fig.~\ref{fig7}a) \\
\hline
\multicolumn{2}{l}{$R$} & $11 \pm 5$ & $13 \pm 7$\\
& $s$ ($s_{\mathrm{opt}}$) & $ 0.61 \pm 0.08$ ($1.02 \pm 0.13$)
& $ 0.60 \pm 0.08$ ($1.00 \pm 0.14$)\\
H$\alpha$ $\biggl\{ \biggr.$ & $F_{\alpha \mathrm{n}}$ &
$19.76 \pm 0.97$ & $19.55 \pm 1.14$\\
& $a$ & $0.53 \pm 0.08$ & $0.11 \pm 0.02$\\
& $s$ ($s_{\mathrm{opt}}$) & $0.85 \pm 0.05$ ($1.41 \pm 0.09$) &
$0.84 \pm 0.06$ ($1.40 \pm 0.10$)\\
H$\beta$ $\biggl\{ \biggr.$ & $F_{\beta \mathrm{n}}$ &
$4.42 \pm 0.22$ & $4.37 \pm 0.26$\\
&$a$ & $0.16 \pm 0.04$ & $0.040 \pm 0.009$\\
\multicolumn{2}{l}{$\chi^2/N$} & 2.97 & 2.98\\
\hline
\end{tabular}

\medskip
$F_{\alpha \mathrm{n}}$ and $F_{\beta \mathrm{n}}$ are in the units of
$10^{-12} \mbox{ erg cm$^{-2}$ s$^{-1}$}$.
\end{center}
\end{table}

\begin{table}
\begin{center}
\caption{The same in the case of $D_\mathrm{n} = 7.55$} \label{tab2}
\begin{tabular}{@{}lccc}
\hline
\multicolumn{2}{c}{} & Orthog.\ orient.\ & Random orient.\ \\
 && (Fig.~\ref{fig6}b) & (Fig.~\ref{fig7}b) \\
\hline
\multicolumn{2}{l}{$R$} & $14 \pm 8$ & $11 \pm 5$\\
&$s$ ($s_{\mathrm{opt}}$) & $0.68 \pm 0.07$
($1.13 \pm 0.12$) & $0.61 \pm 0.10$ ($1.02 \pm 0.16$)\\
H$\alpha$ $\biggl\{ \biggr.$ & $F_{\alpha \mathrm{n}}$ &
$20.96 \pm 0.86$ & $20.22 \pm 1.31$\\
& $a$ & $0.56 \pm 0.13$ & $0.11 \pm 0.01$\\
& $s$ ($s_{\mathrm{opt}}$) & $0.71 \pm 0.05$
($1.18 \pm 0.09$) & $0.66 \pm 0.06$ ($1.10 \pm 0.10$)\\
H$\beta$ $\biggl\{ \biggr.$ & $F_{\beta \mathrm{n}}$ &
$2.78 \pm 0.11$ & $2.68 \pm 0.17$\\
& $a$ & $0.34 \pm 0.09$ & $0.063 \pm 0.009$\\
\multicolumn{2}{l}{$\chi^2/N$} & 3.76 & 3.71 \\
\hline
\end{tabular}
\end{center}
\end{table}

\citet{mushotzky84} elaborated photoionization models of
stationary AGN optical spectra. In the framework of the
photoionization modelling they carried out calculations, in our
equivalent terms representing the calculations of the functions
$L_\mathrm{l}(F_\mathrm{i})$ for the BLR clouds. According to the
results of their calculations, $s \approx 0.6$ for H$\alpha$, and
$s \approx 0.8$ for H$\beta$. Thus, there exists a satisfactory
agreement of the results of our modelling with the data of
\citet{mushotzky84}, especially in the case of $D_\mathrm{n} = 4.47$.
Let us remark that the values of $s$ can be nonconstant inside the
BLR, varying from cloud to cloud, because they depend on physical
characteristics of the clouds. As a result of our modelling we
obtain some ``effective'' values of $s$.

The model of a homogeneous distribution of clouds implies that the
covering factor is close to one, but, as it has been noted above,
an interpretation of the adopted model is possible as a model with
an exponential decrease of the cloud concentration with distance
away from the centre; then the covering factor can be small.

\section[]{Discussion}

There are several ways of estimating the BLR size. Besides the
reverberation and cross-correlation methods, discussed above,
there exists a technique based on estimating the ionization
parameter from modelling the stationary emission-line spectra of
AGN. Using this technique, \citet{mushotzky84} obtained an
estimate of the radius of the BLR of NGC~4151, equal to
approximately 16~lt-days. By a similar argument,
\citet{cassidy97} found the inner and outer radii equal to 6 and
40~lt-days in their theoretical model of the BLR of this galaxy.
At the same time when the photoionization estimate was made by
\citet{mushotzky84}, the reverberation estimate $R \simeq
15$~lt-days was obtained independently by \citet{shevchenko84}, in
agreement with the photoionization estimate by
\citet{mushotzky84}. This reverberation estimation was performed
within the framework of the model of a homogeneous isotropic
distribution of line-emitting matter around the central ionizing
source, on the basis of the observational data of \citet{lyutyi71}
and \citet{cherepashchuk73} on the time lags in the H$\alpha$ line
variations.

Cross-correlation estimates are usually less than the
``photoionization'' values. Cross-correlation analysis by
\cite{peterson88a} (see also discussion by \citealt{peterson88}),
accomplished on the basis of their own observational data and the
data of \citet{antonucci83} on variability in the lines H$\beta$
and \mbox{He\,{\sc ii}} $\lambda 4686$, gave $\sim 6$~lt-days as
the estimate for the radius of the BLR of NGC~4151. Similar
cross-correlation estimates of the BLR size were recovered by
\cite{clavel90} on the basis of the {\it IUE} ({\it International
Ultraviolet Explorer}) data on variability
of the major UV lines: $R = 4 \pm 3$~lt-days. These values
correspond to the peak CCF time lags; the centroid ones are
greater by about two days. \cite{wandel99} find similar centroid
CCF time lags, $4 \pm 3$~d, for the H$\beta$ line. \cite{clavel90}
note that their cross-correlation estimates of the BLR size for
NGC~4151 are an order of magnitude less than the typical
``photoionization'' estimates for Seyfert galaxies.

By means of cross-correlation analysis of their own data,
\citet{kaspi96} found that the time lag of variations in the
H$\alpha$ and H$\beta$ lines in relation to continuum is 0--3~d;
thus the cross-correlation estimate of the BLR radius is 0--3~lt-days.
According to the modern analysis of these data
accomplished by \citet{metzroth06} and \citet{bentz06}, the value
of the cross-correlation time lag for the data of \citet{kaspi96}
has no clear-cut statistical bounds.

In total, the cross-correlation estimates of the BLR radius of
NGC~4151 are all in the range of 0--6~lt-days. Direct
reverberation modelling, in comparison with the cross-correlation
analysis, give very different values of $R$ similar to the given
above ``photoionization'' estimates. According to the results of
\citet{maoz91}, who carried out reverberation modelling of light
curves of NGC~4151 in H$\alpha$ and H$\beta$, the weighted-mean
(by the local emission-line luminosity of clouds) BLR radius
$\approx 16$--18~lt-days for the best found model, and the central
cavity radius $R_0 \approx 2$~lt-days. The linear character of the
$L_\mathrm{l}(F_\mathrm{i})$ dependence was assumed, as in practically
all modern research on this subject. \citet{xue98} numerically
recovered the BLR transfer functions on the basis of the data of
\citet{maoz91} and \citet{kaspi96}. They obtained the following
estimates: $R \approx 10$~lt-days, $R_0 \le 1$~lt-day. As
mentioned above, the reverberation estimate $R \simeq 15$~lt-days
was obtained in \citep{shevchenko84}. All these reverberation
estimates are in agreement with our reverberation modelling
results presented in Tables~\ref{tab1} and \ref{tab2}.

So, the known reverberation estimates of the BLR size of NGC~4151
are in agreement with ``photoionization'' estimates, and they all
are much greater than the cross-correlation estimates. The strong
difference between the BLR radii found by reverberation modelling,
on one side, and its estimates following from cross-correlation
analysis, on the other side, (10--18~lt-days versus 0--6~lt-days)
underlines the conditional character of the cross-correlation
estimates. Such a difference is no surprise: the size identified
as the value of the observed time lag can be much (an order of
magnitude) less than the true size of the BLR in lt-days
(Section~\ref{tlccm}). For example, if the cloud aggregate is
uniform, the time lag of variation of a line with $s \approx 1$
with respect to an ionizing flare is small compared to the BLR
radius $R$ in light travel time units, and depends on $R$ only
weakly. The ultimate cause of this phenomenon is the degeneracy of
relation~(\ref{deltat}) at $s \ge 1$. This degeneracy means that
in practice there are no rigourous theoretical grounds to believe
that the $\Delta t_\mathrm{peak}$ value is mostly determined by the
BLR size, if $\Delta t_\mathrm{peak}$ is calculated for a typical
line (i.e., a line with $s \approx 1$).

\begin{figure}
\begin{center}
\includegraphics[width=130mm]{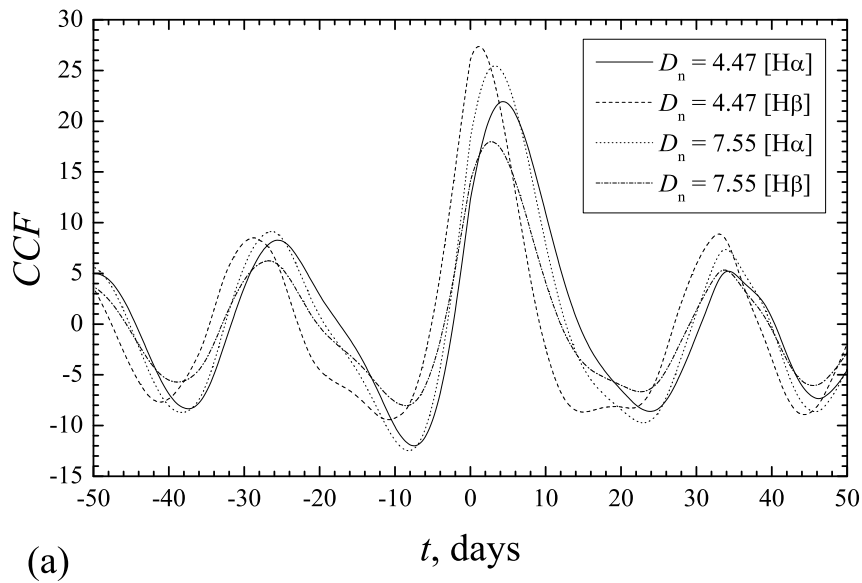}\\
\includegraphics[width=130mm]{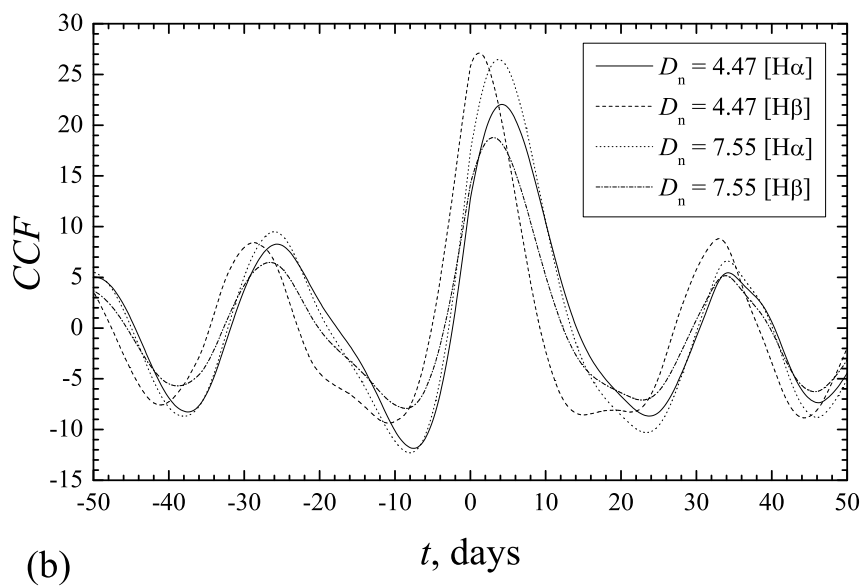}\\
\caption{The computed cross-correlations between the splined curve
in continuum and our theoretical model light curves. The case when
the planes of clouds are orthogonal to the direction to the
central source (a) and the case when they are oriented randomly (b).}
\label{fig11}
\end{center}
\end{figure}

However, cross-correlation analysis by \cite{kaspi96} of their
observational data indicated that the cross-correlation time lag
was small not only for H$\beta$ (the line with $s \approx 1$
presumably), but for H$\alpha$ as well (the line with $s$
definitely less than one). To clarify this point, we have examined
cross-correlations between the splined curve in continuum and our
{\it theoretical} model light curves in H$\alpha$ (presented in
Figs.~\ref{fig6} and \ref{fig7}; linear trends have been
subtracted prior to the analysis). The analysis has shown that the
cross-correlation time lag is 0--3~d, practically equal to the
cross-correlation time lag found by \citet{kaspi96} for the
observed emission-line light curves; the computed
cross-correlation functions are shown in Fig.~\ref{fig11}. So,
substitution of the observed emission-line light curve by a
theoretical one calculated for a definite value of $R$ (equal to
11--14~lt-days) in the transfer function does not change the
cross-correlation estimate of $R$ (0--3~lt-days). It is much less
than the value adopted as the parameter of the transfer function.
This means that the cross-correlation method may provide
inadequate results not only at $s \approx 1$, but at $s < 1$ as
well. The cause is not all-together clear, but one may speculate
that it is related to the strong dependence of $\Delta
t_\mathrm{peak}$ on the variability time scale (Fig.~\ref{fig5})
and/or to the small amplitudes of variability in this particular
set of observational data. Let us underline that, contrary to the
cross-correlation analysis, the reverberation modelling turns out
to be immune, as we have seen, to these unfavourable conditions,
and provides adequate values of $R$.

A hypothesis on possible variability of the BLR size of NGC~4151
was put forward by \citet{kaspi96}, \citet{peterson02} and other
researchers on the basis of cross-correlation analysis of optical
spectral variability data at different time intervals of
observations. Our reverberation modelling of the H$\alpha$ light
curve data of \citet{kaspi96} gives the value of the BLR radius
matching the majority of the BLR size estimates of other authors.
This removes necessity in any special physical interpretation of
the small value of the cross-correlation time lag in H$\alpha$ for
these light curve data. In particular, the hypothesis by
\citet{kaspi96} that the physical size of the BLR at the moment of
their observations was an order of magnitude less than usually is
not necessitated.

\section{Conclusions}

We have studied how the nonlinearity in the
``$L_\mathrm{l}$--$F_\mathrm{i}$'' relation (the emission-line
luminosity, $L_\mathrm{l}$, of the BLR cloud in dependence on the
ionizing continuum flux, $F_\mathrm{i}$, incident on the cloud) can
be taken into account in estimating the size of the BLR in active
galactic nuclei by means of ``reverberation'' methods. We have
shown that the BLR size estimates obtained by cross-correlation
peaks of emission-line and continuum light curves can be much (up
to an order of magnitude) less than the values obtained by
reverberation modelling. This has been demonstrated by means of
abstract representative numerical cross-correlation and
reverberation experiments with model continuum flares and
emission-line transfer functions and by means of practical
reverberation modelling of the observed emission-line variability
of NGC~4151. The modelling of the observed light curves of
NGC~4151 in H$\alpha$ and H$\beta$ has been accomplished on the
basis of the observational data by \citet{kaspi96} and the
theoretical broad-line region model by \citet{shevchenko84,
shevchenko85a}.

In the abstract representative numerical cross-correlation and
reverberation experiments with model continuum flares and
emission-line transfer functions, we have found that the value of
the cross-correlation peak time lag $\Delta t_\mathrm{peak}$ for $s
\ge 1$ is small in comparison with the BLR size $R$ expressed in
the light-travel time units and depends on $R$ only weakly. We
have shown that in the case of  $s \ge 1$ the effect of the
ionizing flare duration on the $\Delta t_\mathrm{peak}$ value is far
greater than that of the BLR radius. In other words, the BLR
radius has little effect on the measured value of the $\Delta
t_\mathrm{peak}$ value in the mathematically degenerate but
observationally most common case of $s=1$; the role of the
timescale of variability is far greater. Therefore, the lines with
$s \approx 1$ seem to be of little help in determining the
size of the BLR by means of estimating the cross-correlation peak
time lag.

The presence of a noticeable time lag of variations of NGC~4151 in
H$\alpha$ \citep{lyutyi71, cherepashchuk73} and significantly
shorter time lags in other Balmer lines \citep{antonucci83} with
respect to variations in the optical continuum has been
attributed, in agreement with conclusions by \citet{shevchenko84,
shevchenko85a}, to the effect of essential nonlinearity in the
``$L_\mathrm{l}$--$F_\mathrm{i}$'' relation for H$\alpha$. The low value
of the power-law index, $s \approx 0.6$, distinguishes this line
from the other Balmer lines.

The values of the model parameters of the BLR of NGC~4151 have
been estimated. In particular, estimates of the BLR radius have
been made. Our reverberation modelling of the emission-line
variability based on the observational data by \citet{kaspi96}
gives values of the BLR radius agreeing with the majority of its
known ``reverberation'' and ``photoionization'' estimates. Much
smaller $R$ values obtained by means of the cross-correlation
method have been shown to be an artifact of this method. The
hypothesis by \citet{kaspi96} that the size of the BLR of NGC~4151
at the time interval of their observations were an order of
magnitude less than usually is not necessitated.

Concluding, a power-law emission-line response model and
simple spherically symmetric thick geometries of the BLR cloud
distribution, taken here as a basis for modelling the
emission-line AGN variability, gives the size of the BLR of
NGC~4151 equal to 11--14~lt-days, with the uncertainty of
5--8~lt-days. This agrees satisfactorily with BLR size estimates
in photoionization models fitting emission line strengths in the
mean spectrum of NGC~4151. Much shorter time lags found in the
cross-correlation analysis of the emission-line and continuum
light curves of NGC~4151 correspond to the size of a smaller
emission-line region that is reverberating.

\section*{Acknowledgments}

The authors thank anonymous referees, whose advice and remarks led
to a significant improvement of the manuscript. We express
appreciation to S.G.Sergeev for extremely valuable comments.
We are deeply grateful to V.V.Kouprianov for programming
assistance and discussions. We thank E.Yu.Aleshkina for technical
help. This work was supported by the Programme of Fundamental
Research of the Russian Academy of Sciences ``Origin and Evolution
of Stars and Galaxies''. A.V.Melnikov is grateful to the Russian
Science Support Foundation for support. The computations were
partially carried out on the computers of the St.Petersburg Branch
of the Supercomputer Centre of the Russian Academy of Sciences.

\label{lastpage}

\end{document}